\documentclass[aps, prx, twocolumn, superscriptaddress]{revtex4-1}
\usepackage{bm, amsmath, amsfonts, amssymb, braket}
\usepackage[dvipdfmx]{graphicx}
\usepackage{float}
\usepackage{color}
\usepackage{comment}
\usepackage{xspace}
\usepackage{stmaryrd}

\newcommand{\ii}{\text{i}}

\begin{document}

\title{Anomalous helical edge states in a non-Hermitian Chern insulator}

\author{Kohei Kawabata}
\email{kawabata@cat.phys.s.u-tokyo.ac.jp}
\affiliation{Department of Physics, University of Tokyo, 7-3-1 Hongo, Bunkyo-ku, Tokyo
113-0033, Japan}

\author{Ken Shiozaki}
%\email{ken.shiozaki@riken.jp}
\affiliation{Condensed Matter Theory Laboratory, RIKEN, Wako, Saitama 351-0198, Japan}
%\affiliation{Yukawa Institute for Theoretical Physics, Kyoto University, Kyoto 606-8502, Japan}

\author{Masahito Ueda}
%\email{ueda@phys.s.u-tokyo.ac.jp}
\affiliation{Department of Physics, University of Tokyo, 7-3-1 Hongo, Bunkyo-ku, Tokyo
113-0033, Japan}
\affiliation{RIKEN Center for Emergent Matter Science (CEMS), Wako, Saitama 351-0198, Japan}

\date{\today}

\begin{abstract}
A non-Hermitian extension of a Chern insulator and its bulk-boundary correspondence are investigated. It is shown that in addition to the robust chiral edge states that reflect the nontrivial topology of the bulk (nonzero Chern number), anomalous helical edge states localized only at one edge can appear, which are unique to the non-Hermitian Chern insulator. 
\end{abstract}

\maketitle

%%%%% Introduction %%%%%
\section{Introduction}

Over the past decade, remarkable progress has been achieved in phases of matter characterized by the topology of their wave functions~\cite{Kane-review, Zhang-review, Schynder-Ryu-review, Bernevig-textbook, Asboth-textbook}. Such topological phases were explored in solid-state systems including insulators~\cite{TKNN-82, Haldane-88, Kane-Mele-05, BHZ-06, Moore-07, Fu-07, Konig-07, Hsieh-08}, superconductors~\cite{Read-00, Kitaev-01, Mourik-12, Alicea-review, Sato-review}, and semimetals~\cite{Wan-11, Burkov-11, Armitage-review}, as well as in photonic~\cite{Haldane-08, Wang-09, Rechtsman-13, Lu-review, Ozawa-review} and atomic~\cite{Atala-13, Jotzu-14, Aidelsburger-15, Nakajima-16, Lohse-16, Goldman-review, Cooper-review} systems, all of which are classified according to spatial dimension and symmetry~\cite{Qi-08, Schnyder-08, Kitaev-09, Ryu-10, Teo-10, Slager-13, Shiozaki-14, Kruthoff-17}. A hallmark of these topological phases is the bulk-boundary correspondence: the topologically protected gapless boundary states appear as a consequence of the nontrivial topology of the gapped bulk. Examples include chiral edge states in Chern insulators~\cite{Haldane-88}, helical edge states in quantum spin Hall insulators~\cite{Kane-Mele-05, BHZ-06, Moore-07, Fu-07}, and Majorana zero modes in topological superconducting wires~\cite{Kitaev-01}.

Recently, there has been growing interest in non-Hermitian topological phases of matter both in theory~\cite{Rudner-09, Hu-11, Esaki-11, Liang-13, Schomerus-13, Malzard-15, SanJose-16, Lee-16, Harter-16, Leykam-17, Xu-17, Menke-17, Gonzalez-17, Saleur-17, Lieu-18, Cerjan-18, Zyuzin-18, Alvarez-18, Xiong-18, Molina-18, Shen-18, Kozii-17, KK-18-TSC, Ni-18, Papaj-18, Gong-18, Yao-18, KK-18-unifies, Yoshida-18, Yin-18, Shen-18-QO} and experiment~\cite{Zeuner-15, Weimann-17, Obuse-17, Zhou-18, Zhao-18, Parto-18, Segev-18, Pan-18}. In general, non-Hermiticity arises from the exchange of energy and/or particles with the environment~\cite{Bender-review, Konotop-review, Feng-review, Christodoulides-review}, and several phenomena unique to the nonconservative systems have been theoretically proposed~\cite{Hatano-96, Bender-98, Bender-07, Musslimani-08, Makris-08, Graefe-08, Gunther-08, Longhi-09, Longhi-10, Chong-11, Lin-11, Szameit-11, Brody-12, Wiersig-14, Lee-14x, Lee-14, Tripathi-16, Liu-16, Ashida-16, Ge-17, Yin-17, KK-PT-17, Ge-18, Konotop-18, Ashida-18, Nori-18, Spielman-18, Lourenco-18, Nakagawa-18} and experimentally observed~\cite{Guo-09, Ruter-10, Regensburger-12, Feng-13, Peng-14, Feng-14, Hodaei-14, Zhen-15, Gao-15, Tang-16, Peng-16, Hodaei-17, Chen-17}. A key feature of non-Hermitian systems is the presence of a level degeneracy called an exceptional point~\cite{Berry-04, Heiss-12, Moiseyev-11}, at which eigenstates coalesce to render the Hamiltonian nondiagonalizable. The exceptional point brings about novel functionalities with no Hermitian counterparts such as unidirectional invisibility~\cite{Lin-11, Regensburger-12, Feng-13, Peng-14} and enhanced sensitivity~\cite{Wiersig-14, Liu-16, Hodaei-17, Chen-17}. Recent studies have also revealed that non-Hermiticity alters the nature of the bulk-boundary correspondence in topological systems~\cite{Hu-11, Schomerus-13, Malzard-15, Lee-16, Leykam-17, Alvarez-18, Xiong-18, Yao-18}. Non-Hermiticity was shown to amplify the topologically protected edge states~\cite{Schomerus-13}, which were experimentally observed in one dimension~\cite{Obuse-17, Parto-18} and two dimensions~\cite{Segev-18}. Furthermore, the presence of exceptional points makes edge states anomalous, so that they are localized only at one edge in a non-Hermitian extension of the Su-Schrieffer-Heeger model (i.e., a non-Hermitian system in one dimension that respects sublattice symmetry)~\cite{Lee-16}. Despite these recent studies, the bulk-boundary correspondence in non-Hermitian systems has yet to be fully understood, especially in two dimensions.

In this work, we investigate a non-Hermitian extension of a Chern insulator and its bulk-boundary correspondence. In Sec.~\ref{sec: topo inv}, we give the topological invariants for a general non-Hermitian system in two dimensions without symmetry (2D class A). In Sec.~\ref{sec: phase}, we consider a typical non-Hermitian Chern insulator and provide its phase diagram under the periodic boundary condition. It consists of the gapped phases characterized by the Chern number and the gapless phases characterized by the topological charges accompanied by exceptional points. In Sec.~\ref{sec: edge}, we investigate the topologically protected edge states. The chiral edge states are shown to be robust even in the non-Hermitian systems; they can also be amplified (lasing) due to the presence of non-Hermiticity. Moreover, we find the emergence of the helical edge states localized only at one edge in some phases. These anomalous helical edge states do not have Hermitian counterparts and hence they are unique to the non-Hermitian Chern insulator. In Sec.~\ref{sec: conclusion}, we conclude the paper with an outlook.

%%%%% 2D class A %%%%%
\section{Topological invariants}
	\label{sec: topo inv}

We first provide the topological invariants of a general non-Hermitian system in two dimensions without symmetry (2D class A). If the Hamiltonian $H \left( {\bm k} \right)$ is diagonalizable, it can be expressed as
\begin{equation}
H \left( {\bm k} \right)
= \sum_{n} E_{n} \left( {\bm k} \right) \ket{\varphi_{n} \left( {\bm k} \right)} \bra{\chi_{n} \left( {\bm k} \right)}.
\end{equation}
Here $E_{n} \left( {\bm k} \right)$ is a complex eigenenergy and $\ket{\varphi_{n} \left( {\bm k} \right)}$ ($\ket{\chi_{n} \left( {\bm k} \right)}$) is the corresponding right (left) eigenstate~\cite{Brody-14, Brody-16}, which satisfy
\begin{equation} \begin{split}
H \left( {\bm k} \right) \ket{\varphi_{n} \left( {\bm k} \right)}
&= E_{n} \left( {\bm k} \right) \ket{\varphi_{n} \left( {\bm k} \right)}, \\
H^{\dag} \left( {\bm k} \right) \ket{\chi_{n} \left( {\bm k} \right)}
&= E_{n}^{*} \left( {\bm k} \right) \ket{\chi_{n} \left( {\bm k} \right)},
\end{split} \end{equation}
and 
\begin{equation} \begin{split}
\braket{\chi_{m} \left( {\bm k} \right) | \varphi_{n} \left( {\bm k} \right)} 
&= \delta_{mn}, \\
\sum_{n} \ket{\varphi_{n} \left( {\bm k} \right)} \bra{\chi_{n} \left( {\bm k} \right)}
&= I,
\end{split} \end{equation}
with the identity matrix $I$. Here $\ket{\varphi_{n} \left( {\bm k} \right)}$ and $\ket{\chi_{n} \left( {\bm k} \right)}$ are equal to each other in the presence of Hermiticity, but they are not for general non-Hermitian Hamiltonians. If the complex band $n$ is separated from the others (i.e., $\forall\,{\bm k},\,m~~E_{m} \left( {\bm k} \right) \neq E_{n} \left( {\bm k} \right)$), we can define the Berry connection ${\cal A}_{n}^{i} \left( {\bm k} \right)$ and the Berry curvature ${\cal F}_{n} \left( {\bm k} \right) $ of the complex band $n$ as
\begin{eqnarray}
{\cal A}_{n}^{i} \left( {\bm k} \right) &:=& \ii \braket{\chi_{n} \left( {\bm k} \right) |\,\partial_{k_{i}} \varphi_{n} \left( {\bm k} \right)}, \\
{\cal F}_{n} \left( {\bm k} \right) &:=& \partial_{k_{x}} {\cal A}_{n}^{y} \left( {\bm k} \right) - \partial_{k_{y}} {\cal A}_{n}^{x} \left( {\bm k} \right),
\end{eqnarray}
and the (first) Chern number as~\cite{Liang-13, Shen-18}
\begin{equation}
C_{n} := \int_{\rm BZ} \frac{d^{2}{\bm k}}{2\pi} {\cal F}_{n}
\in \mathbb{Z},
	\label{eq: Chern - def}
\end{equation}
where the integration is taken over the first Brillouin zone (BZ). The Chern number remains unchanged as long as the complex band is separated and the Hamiltonian is diagonalizable. Due to the difference between the left and right eigenstates, there is some arbitrariness concerning the definition of the Berry connection. For instance, it can also be defined only by the right eigenstates as $\tilde{\cal A}_{n}^{i} \left( {\bm k} \right) := \ii \braket{\varphi_{n} \left( {\bm k} \right) |\,\partial_{k_{i}} \varphi_{n} \left( {\bm k} \right)}$. However,  the Chern number does not depend on such choices of the eigenstates and can be uniquely defined~\cite{Shen-18}.

On the other hand, the Chern number is not well-defined if the Hamiltonian is nondiagonalizable at a wavenumber ${\bm k} = {\bm k_{\rm EP}}$ (i.e., some eigenstates coalesce and linearly depend on each other)~\cite{Berry-04, Heiss-12, Moiseyev-11}. Instead, in the presence of such a defective point (exceptional point), we can define another topological invariant given by~\cite{Leykam-17, Xu-17, Zhou-18, Shen-18, Zhen-15, Gao-15} 
\begin{equation}
\nu \left( {\bm k_{\rm EP}} \right)
:= \oint_{\Gamma \left( {\bm k_{\rm EP}} \right)} \frac{d{\bm k}}{2\pi \ii} \cdot \nabla_{\bm k} \log \left( E_{m} - E_{n} \right),
	\label{eq: EP - charge}
\end{equation}
where the two bands $E_{m} \left( {\bm k} \right)$ and $E_{n} \left( {\bm k} \right)$ coalesce at ${\bm k} = {\bm k_{\rm EP}}$, and $\Gamma \left( {\bm k_{\rm EP}} \right)$ is a closed loop in momentum space that encircles ${\bm k_{\rm EP}}$. Here fractional $\nu \left( {\bm k_{\rm EP}} \right)$ implies that $E_{m} \left( {\bm k} \right) - E_{n} \left( {\bm k} \right)$ is multi-valued and that ${\bm k_{\rm EP}}$ becomes a branch point in momentum space. The charge $\nu \left( {\bm k_{\rm EP}} \right)$ defined as Eq.~(\ref{eq: EP - charge}) is topological in that it cannot be changed unless the exceptional point is annihilated.

%%%%% Phases and complex spectra %%%%%
\section{Phases and complex spectra}
	\label{sec: phase}
	
\begin{figure}[t]
\centering
\includegraphics[width=86mm]{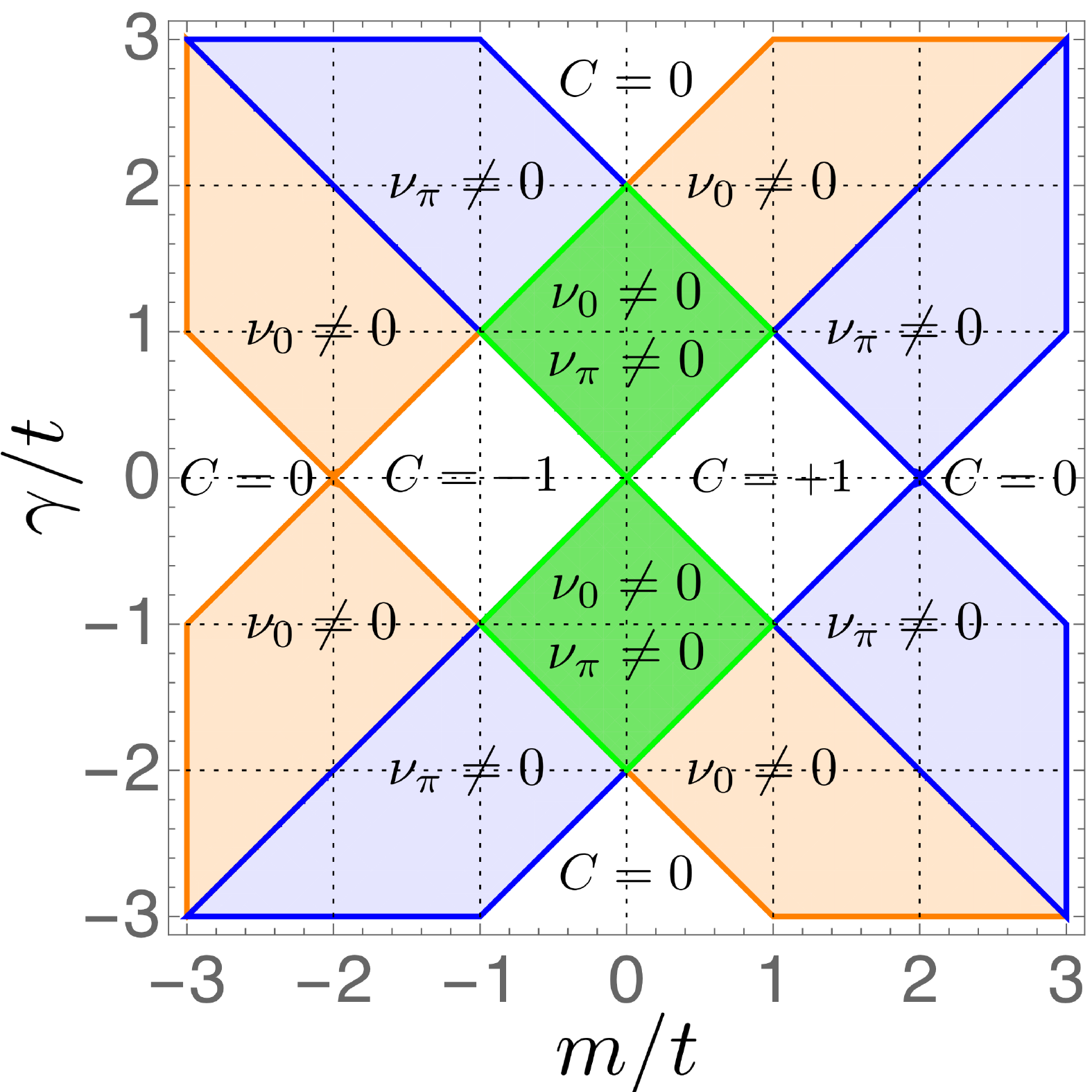} 
\caption{Phase diagram of the non-Hermitian Chern insulator with periodic boundaries given by Eq.~(\ref{eq: Chern insulator}). The white regions represent the gapped phases that are characterized by the Chern number $C$. The orange ($\nu_{0} \neq 0$), blue ($\nu_{\pi} \neq 0$), and green ($\nu_{0} \neq 0,\,\nu_{\pi} \neq 0$) regions represent the gapless phases where pairs of exceptional points appear on $k_{x}=0$, $k_{x} = \pm \pi$, and both $k_{x}=0$ and $k_{x} = \pm \pi$, respectively.}
	\label{fig: Chern-phase}
\end{figure}

\begin{figure}[t]
\centering
\includegraphics[width=85mm]{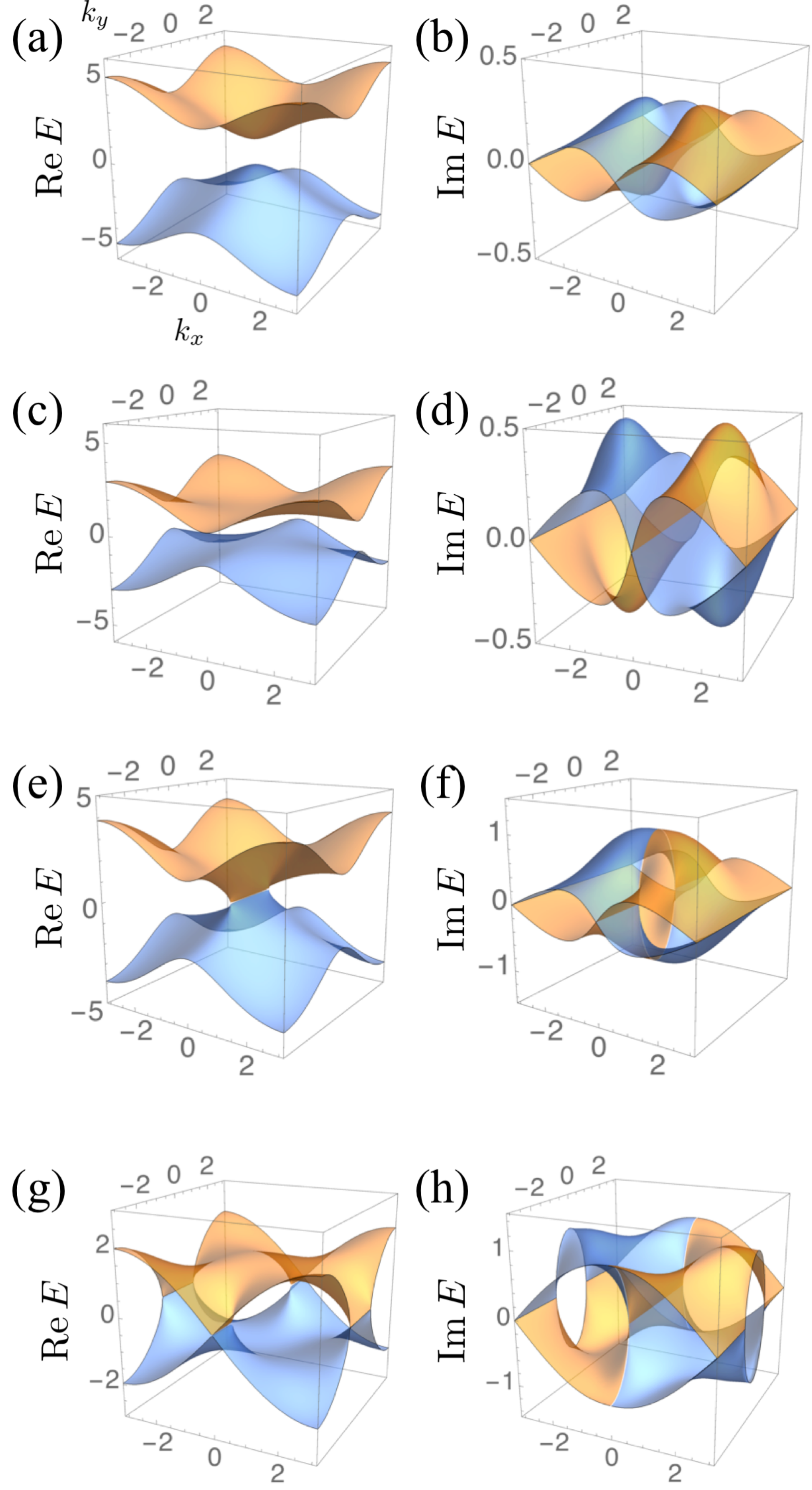} 
\caption{Complex-band structures of the non-Hermitian Chern insulator $E_{\pm} = E_{\pm} \left( k_{x}, k_{y} \right)$ given by Eq.~(\ref{eq: Chern insulator - band}). The orange (blue) band represents $E_{+}$ ($E_{-}$). (a-b)~Real and imaginary parts of the gapped and topologically trivial bands with zero Chern number $C=0$ ($t=1.0,\,m=-3.0,\,\gamma=0.5$). (c-d)~Real and imaginary parts of the gapped and topologically nontrivial bands with nonzero Chern number $C=-1$ ($t=1.0,\,m=-1.0,\,\gamma=0.5$). (e-f)~Real and imaginary parts of the gapless bands with a pair of exceptional points on $k_{x} = 0$ ($t=1.0,\,m=-2.0,\,\gamma=1.0$). (g-h)~Real and imaginary parts of the gapless bands with two pairs of exceptional points on both $k_{x} = 0$ and $k_{x} = \pm \pi$ ($t=1.0,\,m=-0.2,\,\gamma=1.0$).}
	\label{fig: Chern-band}
\end{figure}

In the following, we consider the non-Hermitian Chern insulator given by
\begin{equation} \begin{split}
H \left( {\bm k} \right)
&= \left( m + t \cos k_{x} + t \cos k_{y} \right) \sigma_{x} \\
&~~~~~~~~~+ \left( \ii \gamma + t \sin k_{x} \right) \sigma_{y}
+ \left( t \sin k_{y} \right) \sigma_{z},
	\label{eq: Chern insulator}
\end{split} \end{equation}
where $t$, $m$, and $\gamma$ are real parameters and we assume $t > 0$. In the case of $\gamma = 0$, the model reduces to the well-known Hermitian Chern insulator that is characterized by the Chern number~\cite{Kane-review, Zhang-review, Schynder-Ryu-review, Bernevig-textbook, Asboth-textbook}. The eigenstates form two bands and their complex energy dispersions are obtained as 
\begin{equation} \begin{split}
E_{\pm} \left( {\bm k} \right)
&= \pm \left[ \left( m + t \cos k_{x} + t \cos k_{y} \right)^{2} \right. \\
&~~~~~~~~~+ \left. \left( \ii \gamma + t \sin k_{x} \right)^{2} + \left( t \sin k_{y} \right)^{2} \right]^{1/2}.
	\label{eq: Chern insulator - band}
\end{split} \end{equation}
The Hamiltonian becomes defective if and only if 
\begin{equation}
\exists~{\bm k_{\rm EP}} \in \left[ -\pi,\,\pi \right]^{2}~~s.t.~~E_{\pm} \left( {\bm k_{\rm EP}} \right) = 0.
	\label{eq: EP}
\end{equation}
This requires $\sin \left( k_{\rm EP} \right)_{x} = 0$ and reduces to the following conditions:

\begin{itemize}
\item[(1)] Exceptional points appear on $\left( k_{\rm EP} \right)_{x} = 0$; Eq.~(\ref{eq: EP}) reduces to
\begin{equation}
~~~~~~~~~\gamma^{2}
= \left( m+t\right)^{2} + t^{2} + 2t \left( m+t \right) \cos \left( k_{\rm EP} \right)_{y},
\end{equation}
where $\left( k_{\rm EP} \right)_{y} \in \left[ -\pi,\,\pi \right]$ exists if and only if 
\begin{equation} \begin{cases}
\left| m \right| \leq \left| \gamma \right| \leq \left| m+2t \right|
& {\rm for}~~m \geq -t; \\
\left| m+2t \right| \leq \left| \gamma \right| \leq \left| m \right|
& {\rm for}~~m \leq -t.
\end{cases} \end{equation}

\item[(2)] Exceptional points appear on $\left( k_{\rm EP} \right)_{x} = \pm \pi$; Eq.~(\ref{eq: EP}) reduces to
\begin{equation}
~~~~~~~~~\gamma^{2}
= \left( m-t\right)^{2} + t^{2} + 2t \left( m-t \right) \cos \left( k_{\rm EP} \right)_{y},
\end{equation}
where $\left( k_{\rm EP} \right)_{y} \in \left[ -\pi,\,\pi \right]$ exists if and only if 
\begin{equation} \begin{cases}
\left| m-2t \right| \leq \left| \gamma \right| \leq \left| m \right|
& {\rm for}~~m \geq t; \\
\left| m \right| \leq \left| \gamma \right| \leq \left| m-2t \right|
& {\rm for}~~m \leq t.
\end{cases} \end{equation}
\end{itemize}

Hence we obtain the phase diagram (Fig.~\ref{fig: Chern-phase}) and the corresponding complex band structures (Fig.~\ref{fig: Chern-band}). The gapped phases with $E_{+} \left( {\bm k} \right) \neq E_{-} \left( {\bm k} \right)$ for all ${\bm k}$ (Fig.~\ref{fig: Chern-band}\,(a-d)) are characterized by the Chern number defined by Eq.~(\ref{eq: Chern - def}). Similar to the Hermitian counterparts, the Chern number determines the presence or absence of the chiral edge states, as discussed in the next section. Remarkably, the gapped phases are separated by the broad gapless phases that accompany pairs of exceptional points (Fig.~\ref{fig: Chern-band}\,(e-h)), whereas they are separated by a point in the Hermitian counterpart ($\gamma = 0$). The appearance of such large gapless phases arises from the fact that a level degeneracy requires fine-tuning two parameters for a general $2 \times 2$ non-Hermitian matrix, although it requires fine-tuning three parameters for a general $2 \times 2$ Hermitian matrix~\cite{Shen-18, Yoshida-18, Zhou-18, Berry-04}; non-Hermitian systems in two dimensions can be gapless without fine-tuning parameters or the presence of certain symmetry. As explained in the last section, the topological invariants can be defined for each exceptional point as Eq.~(\ref{eq: EP - charge}). For instance, if there appears a pair of exceptional points at ${\bm k_{\rm EP}} = \left( 0,\,\pm \kappa \right)$ ($\kappa > 0$) as in Fig.~\ref{fig: Chern-band}\,(e-f), the energy dispersions near the exceptional points are given by
\begin{equation}
E_{+} \left( {\bm k} \right) \simeq \sqrt{2\ii t \gamma k_{x} \pm t^{2} \left( \sin 2\kappa \right) \left( k_{y} \mp \kappa \right)}.
\end{equation}
Here the square-root singularity implies that the exceptional points ${\bm k} = {\bm k_{\rm EP}}$ become branch points and that the topological charges defined by Eq.~(\ref{eq: EP - charge}) are $\nu = \pm1/2$ for ${\bm k_{\rm EP}} = \left( 0,\,\pm \kappa \right)$. Although the Chern number cannot be defined, we find the emergence of the anomalous helical edge states in these gapless phases, as discussed in the next section.

%%%%% edge state %%%%%
\section{Edge states}
	\label{sec: edge}

The hallmark of the Hermitian Chern insulator is the emergence of the chiral edge states that correspond to the nontrivial topology of the bulk~\cite{Kane-review, Zhang-review, Schynder-Ryu-review, Bernevig-textbook, Asboth-textbook}. It is shown that the bulk-edge correspondence in the presence of non-Hermiticity is sensitive to the boundary conditions~\cite{remark-BC}. In the following, we investigate the system with open boundaries only in the $y$ direction, the system with open boundaries only in the $x$ direction, and the system with open boundaries in both $x$ and $y$ directions.

\subsection{Open boundaries in the $y$ direction}

\begin{figure}[t]
\centering
\includegraphics[width=86mm]{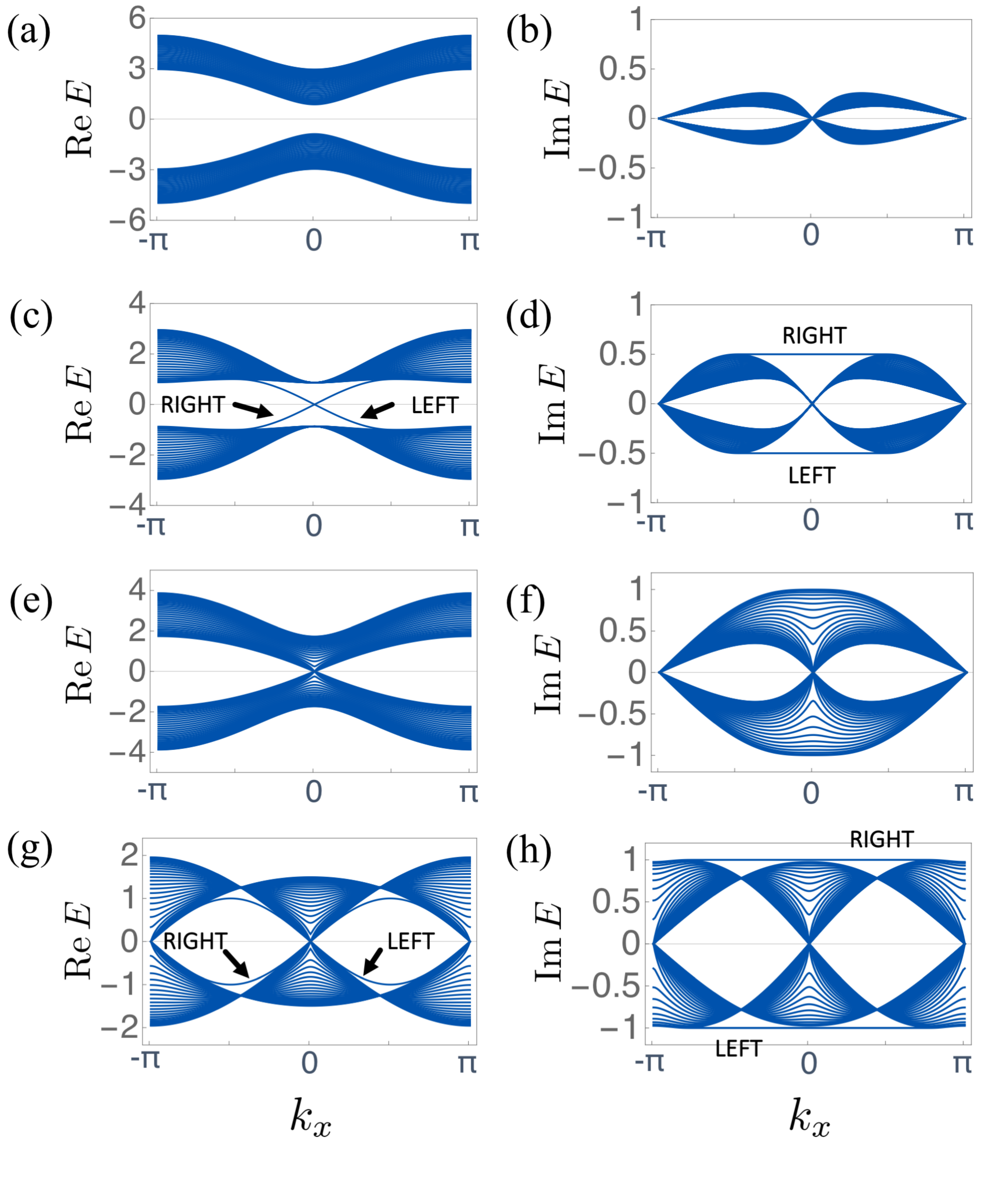} 
\caption{Complex spectrum of the non-Hermitian Chern insulator given by Eq.~(\ref{eq: Chern insulator - yOBC}), which has periodic boundaries in the $x$ direction and open boundaries in the $y$ direction ($L_{y} = 30$). (a-b)~Real and imaginary parts of the complex spectrum in the gapped and topologically trivial phase ($t=1.0,\,m=-3.0,\,\gamma=0.5$). No gapless states appear between the gapped complex bands. (c-d)~Real and imaginary parts of the complex spectrum in the gapped and topologically nontrivial phase ($t=1.0,\,m=-1.0,\,\gamma=0.5$). A pair of chiral edge states appears at both edges between the gapped complex bands whose dispersions are $E_{\rm edge} \left( k_{x} \right) = \pm \left( t \sin k_{x} + \ii \gamma \right)$. (e-f)~Real and imaginary parts of the complex spectrum in the gapless phase with an exceptional point at $k_{x} = 0$ ($t=1.0,\,m=-2.0,\,\gamma=1.0$). (g-h)~Real and imaginary parts of the complex spectrum in the gapless phase with exceptional points at both $k_{x} = 0$ and $k_{x} = \pm \pi$ ($t=1.0,\,m=-0.2,\,\gamma=1.0$). A pair of chiral edge states appears at both edges, despite the absence of the energy gap.}
	\label{fig: Chern-edge-yOBC}
\end{figure}

\begin{figure}[t]
\centering
\includegraphics[width=54mm]{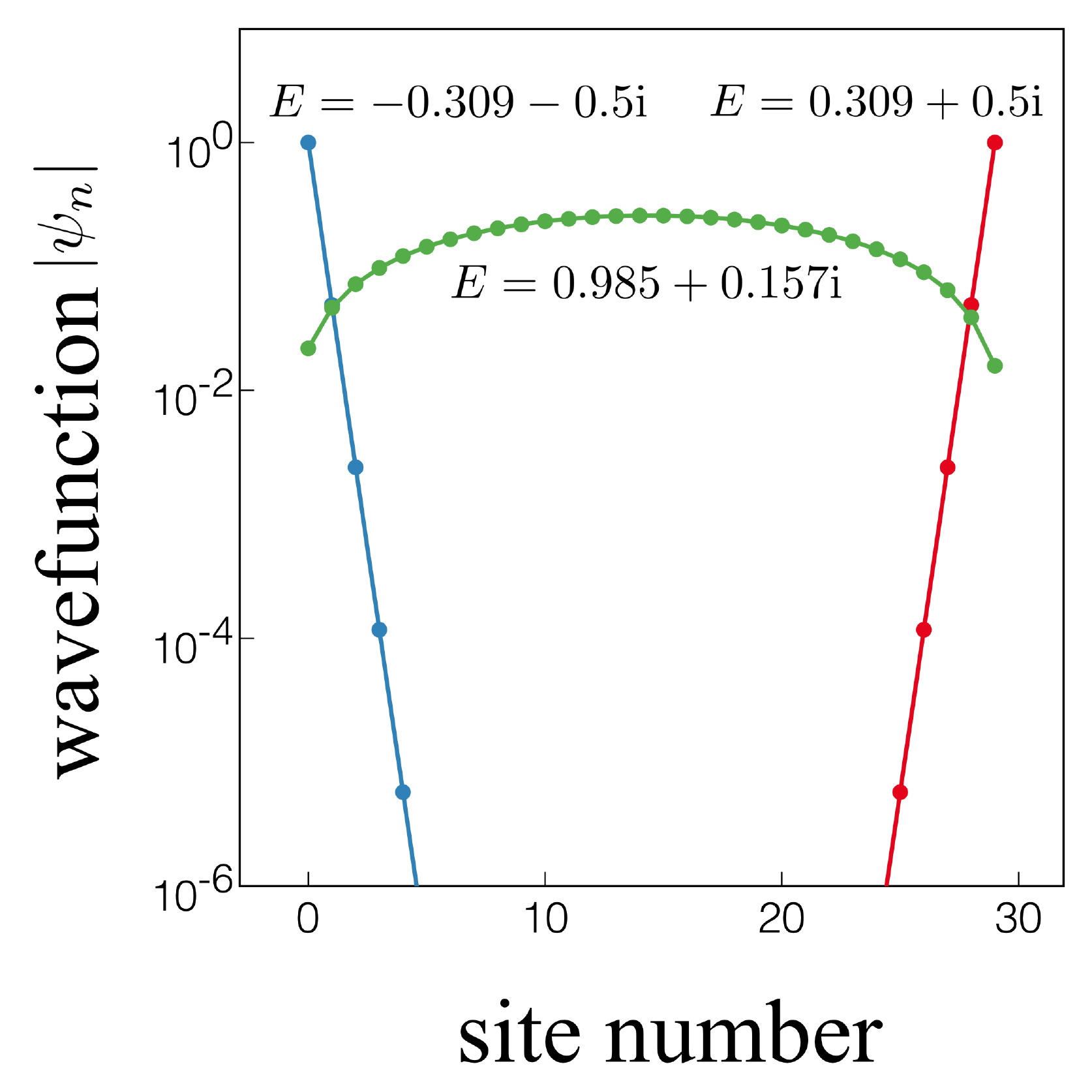} 
\caption{Wavefunctions of the non-Hermitian Chern insulator given by Eq.~(\ref{eq: Chern insulator - yOBC}), which has periodic boundaries in the $x$ direction and open boundaries in the $y$ direction ($L_{y} = 30$). The system is in the gapped and topologically nontrivial phase that corresponds to Fig.~\ref{fig: Chern-edge-yOBC}\,(c-d) ($t=1.0,\,m=-1.0,\,\gamma=0.5;\,k_{x} = 0.1\pi$). Whereas the bulk state with $E = 0.985 + 0.157\ii$ (green dots) is delocalized throughout the bulk, the chiral edge states with $E = \pm \left( 0.309 + 0.5\ii \right)$ (red and blue dots) are localized at both edges.}
	\label{fig: Chern-edge-yOBC-eigenvector}
\end{figure}

We first consider the following non-Hermitian Chern insulator with periodic boundaries in the $x$ direction and open boundaries in the $y$ direction:
\begin{equation} \begin{split}
&\hat{H} = \sum_{k_{x}} \sum_{y} \left\{ \left[ \hat{c}_{k_{x}, y+1}^{\dag} \frac{t \left( \sigma_{x} + \ii \sigma_{z} \right)}{2} \hat{c}_{k_{x}, y} + {\rm H.c.} \right] \right. \\
&+ \left. \hat{c}_{k_{x}, y}^{\dag} \left[ \left( m + t \cos k_{x} \right) \sigma_{x} + \left( \ii \gamma + t \sin k_{x} \right) \sigma_{y} \right] \hat{c}_{k_{x}, y} \right\},
		\label{eq: Chern insulator - yOBC}
\end{split} \end{equation}
where $\hat{c}_{k_{x}, y}$ ($\hat{c}_{k_{x}, y}^{\dag}$) annihilates (creates) a fermion with two internal degrees of freedom on site $y$ and with momentum $k_{x}$.

In the gapped phase with the topologically trivial bulk (Fig.~\ref{fig: Chern-band}\,(a-b)), no gapless states appear between the gapped complex bands (Fig.~\ref{fig: Chern-edge-yOBC}\,(a-b)). In the gapped phase with the topologically nontrivial bulk (Fig.~\ref{fig: Chern-band}\,(c-d)), on the other hand, chiral edge states appear, one localized at the right edge and the other localized at the left edge (Fig.~\ref{fig: Chern-edge-yOBC}\,(c-d) and Fig.~\ref{fig: Chern-edge-yOBC-eigenvector}). The energy dispersions of these chiral edge states are analytically obtained as
\begin{equation} \begin{split}
E_{\rm left} \left( k_{x} \right)
= - t \sin k_{x} - \ii \gamma,~~
E_{\rm right} \left( k_{x} \right)
= t \sin k_{x} + \ii \gamma,
	\label{eq: chiral edge - energy}
\end{split} \end{equation}
and the corresponding edge states are given by
\begin{equation} \begin{split}
\hat{\Psi}_{\rm left} \left( k_{x} \right)
&\propto \sum_{y=1}^{L_{y}} \left( - \cos k_{x} - \frac{m}{t} \right)^{y-1} \hat{c}_{k_{x}, y} \begin{pmatrix} 1 \\ -\ii \end{pmatrix}, \\
\hat{\Psi}_{\rm right} \left( k_{x} \right)
&\propto \sum_{y=1}^{L_{y}} \left( - \cos k_{x} - \frac{m}{t} \right)^{y-1} \hat{c}_{k_{x}, y} \begin{pmatrix} 1 \\ \ii \end{pmatrix},
	\label{eq: chiral edge - state}
\end{split} \end{equation}
which satisfy $[ \hat{H},\,\hat{\Psi}_{\rm left/right} ] = E_{\rm left/right}\,\hat{\Psi}_{\rm left/right}$ in the thermodynamic limit $L_{y} \to \infty$ (see Appendix~\ref{sec: chiral edge states} for the detailed derivation~\cite{remark-calc, Mong-11, Viola-16, Kawabata-17-Majorana}). Here $\left| \cos k_{x} + m/t \right| < 1$ is required for the localization (normalization) of these edge states. The obtained edge states are immune to disorder and hence topologically protected (see Appendix~\ref{sec: disorder} for details). Remarkably, the right edge state has the largest imaginary part for $\gamma > 0$, whereas the left edge state has the smallest imaginary part. Physically, this results in the amplification (lasing) of the right edge state and the attenuation of the left edge state, just like a topological insulator laser recently proposed and realized~\cite{Segev-18}.

In the gapless phases, there appear exceptional points on $k_{x} = 0$ and/or $k_{x} = \pm \pi$ (Fig.~\ref{fig: Chern-edge-yOBC}\,(e-h)). Whereas no edge states appear in Fig.~\ref{fig: Chern-edge-yOBC}\,(e-f), a pair of chiral edge states emerges at both edges as shown in Fig.~\ref{fig: Chern-edge-yOBC}\,(g-h), which are also described by Eqs.~(\ref{eq: chiral edge - energy}) and (\ref{eq: chiral edge - state}). The latter case corresponds to the gapless phase with $\nu_{0},\,\nu_{\pi} \neq 0$ between the gapped $C = +1$ and $C=-1$ phases (see the green regions in Fig.~\ref{fig: Chern-phase}). Such a gapless phase connects the gapped and topologically nontrivial phases, provided that $C_{4}$ rotational symmetry is maintained. Despite the absence of the energy gap, these chiral edge states are also stable against weak perturbations (see Appendix~\ref{sec: disorder} for details), which implies their topological origin. We note that Ref.~\cite{Sato-10} discusses a similar gapless topological phase in Hermitian topological superconductors in two dimensions.

%%%%%
\subsection{Open boundaries in the $x$ direction}

\begin{figure}[t]
\centering
\includegraphics[width=86mm]{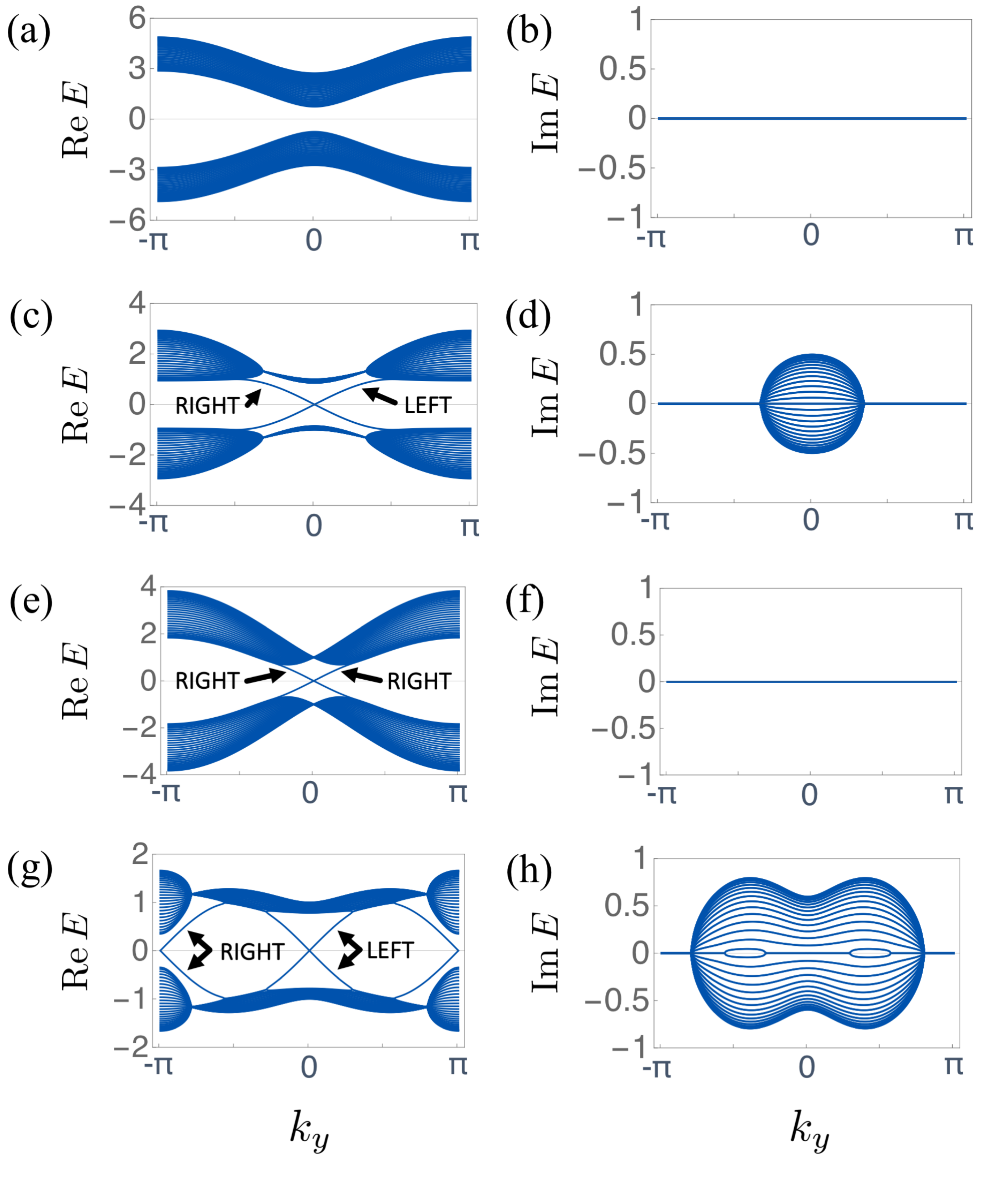} 
\caption{Complex spectrum of the non-Hermitian Chern insulator given by Eq.~(\ref{eq: Chern insulator - xOBC}), which has open boundaries in the $x$ direction ($L_{x} = 30$) and periodic boundaries in the $y$ direction. (a-b)~Real and imaginary parts of the complex spectrum in the gapped and topologically trivial phase ($t=1.0,\,m=-3.0,\,\gamma=0.5$). No gapless states appear between the gapped complex bands. (c-d)~Real and imaginary parts of the complex spectrum in the gapped and topologically nontrivial phase ($t=1.0,\,m=-1.0,\,\gamma=0.5$). A pair of chiral edge states appears between the gapped complex bands. (e-f)~Real and imaginary parts of the complex spectrum for $t=1.0$, $m=-2.0$, and $\gamma=1.0$. Although the complex bands are gapless for the corresponding periodic systems (Fig.~\ref{fig: Chern-band}\,(e-f)), the gap is open and the helical edge states localized only at the right edge appear. (g-h)~Real and imaginary parts of the complex spectrum for $t=1.0$, $m=-0.2$, and $\gamma=1.0$. Although the complex bands are gapless for the corresponding periodic systems (Fig.~\ref{fig: Chern-band}\,(g-h)), the gap is open and two pairs of helical edge states appear, one localized at the right edge and the other localized at the left edge.}
	\label{fig: Chern-edge-xOBC}
\end{figure}

\begin{figure}[t]
\centering
\includegraphics[width=86mm]{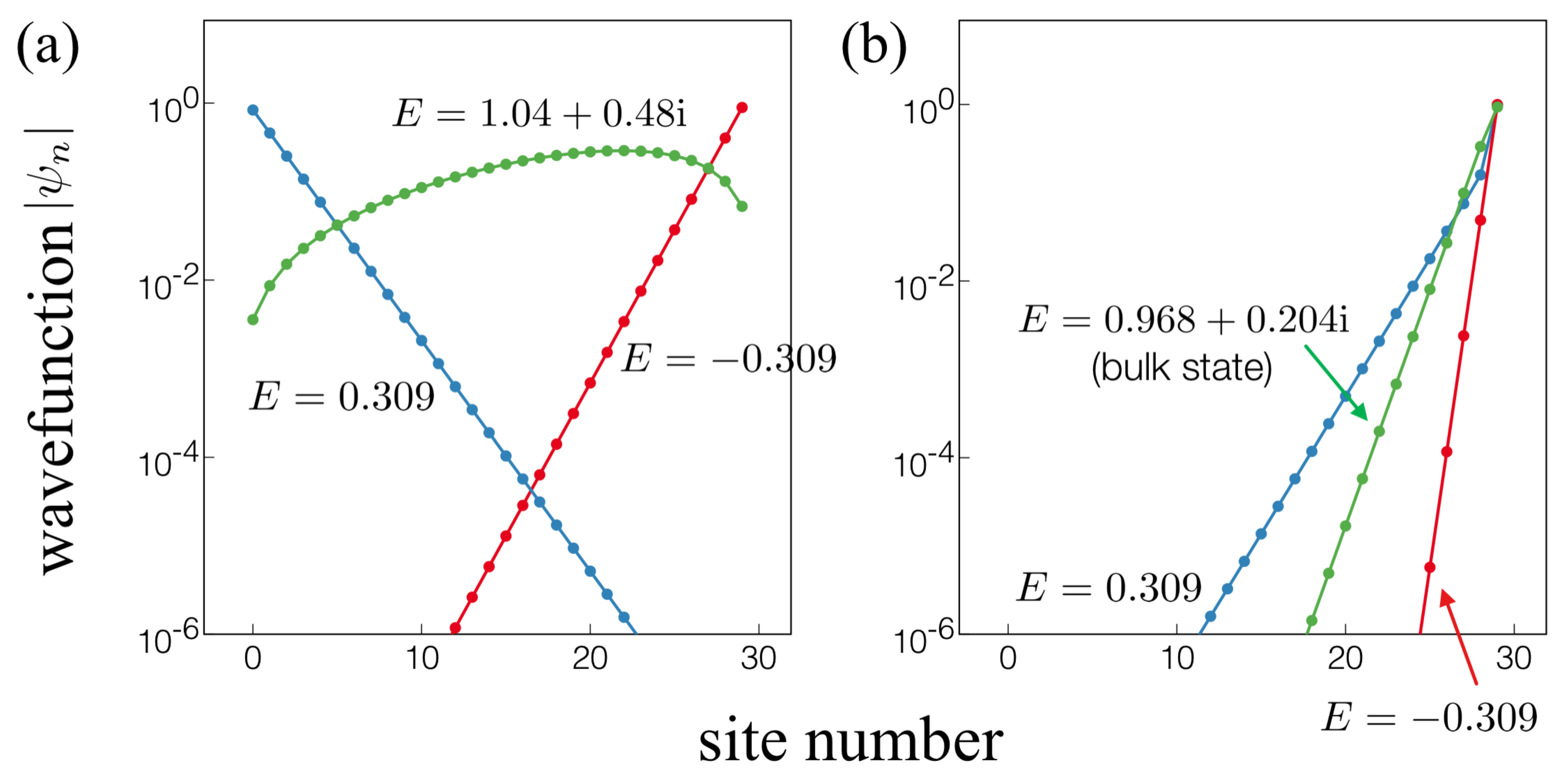} 
\caption{Wavefunctions of the non-Hermitian Chern insulator given by Eq.~(\ref{eq: Chern insulator - xOBC}), which has open boundaries in the $x$ direction ($L_{x} = 30$) and periodic boundaries in the $y$ direction. (a)~The system is in the gapped and topologically nontrivial phase that corresponds to Fig.~\ref{fig: Chern-edge-xOBC}\,(c-d) ($t=1.0,\,m=-1.0,\,\gamma=0.5;\,k_{x} = 0.1\pi$). The chiral edge states with $E = \pm 0.309$ (red and blue dots) at both edges are more localized than the bulk state with $E=1.04+0.48\ii$ (green dots). (b)~The system is in the gapped phase that corresponds to Fig.~\ref{fig: Chern-edge-xOBC}\,(e-f) ($t=1.0,\,m=-2.0,\,\gamma=1.0;\,k_{x} = 0.1\pi$). The helical edge states with $E =\pm 0.309$ (red and blue dots) are localized only at the right edge, one of which is more localized than the bulk state with $E = 0.968 + 0.204\ii$ (green dots) and the other of which is not.}
	\label{fig: Chern-edge-xOBC-eigenvector}
\end{figure}

\begin{figure}[t]
\centering
\includegraphics[width=86mm]{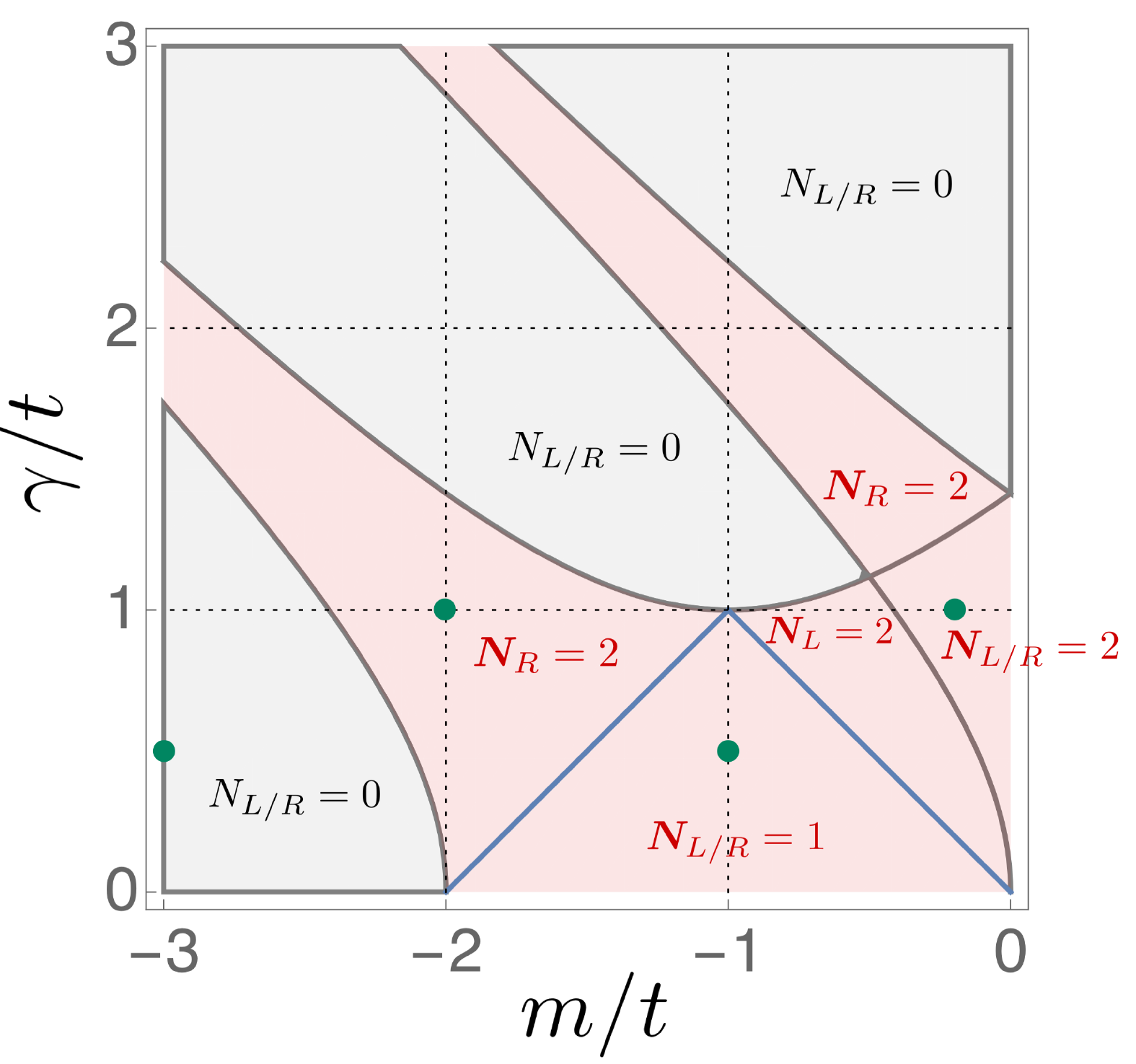} 
\caption{Phase diagram of the non-Hermitian Chern insulator  given by Eq.~(\ref{eq: Chern insulator - xOBC}), which has open boundaries in the $x$ direction and periodic boundaries in the $y$ direction. The gapped phases are distinguished by the presence (red regions) or absence (gray regions) of the edge states, and $N_{L}$ ($N_{R}$) is the number of them localized at the left (right) edge. Across the blue lines, the number of edge states localized at either left or right edge changes, whereas the total number of edge states does not. The four green dots represent the set of parameters used in Fig.~\ref{fig: Chern-edge-xOBC}. In stark contrast to the Hermitian counterparts, the phase diagram is drastically different from that under the periodic boundary condition (Fig.~\ref{fig: Chern-phase}).}
	\label{fig: Chern-xOBC-phase}
\end{figure}

We next consider the following non-Hermitian Chern insulator with open boundaries in the $x$ direction and periodic boundaries in the $y$ direction:
\begin{equation} \begin{split}
&\hat{H} = \sum_{x} \sum_{k_{y}} \left\{ \left[ \hat{c}_{x+1, k_{y}}^{\dag} \frac{t \left( \sigma_{x} + \ii \sigma_{y} \right)}{2} \hat{c}_{x, k_{y}} + {\rm H.c.} \right] \right. \\
&+ \left. \hat{c}_{x, k_{y}}^{\dag} \left[ \left( m + t \cos k_{y} \right) \sigma_{x} + \ii \gamma\,\sigma_{y} + \left( t \sin k_{y}\right) \sigma_{z} \right] \hat{c}_{x, k_{y}} \right\},
	\label{eq: Chern insulator - xOBC}
\end{split} \end{equation}
where $\hat{c}_{x, k_{y}}$ ($\hat{c}_{x, k_{y}}^{\dag}$) annihilates (creates) a fermion with two internal degrees of freedom on site $x$ and with momentum $k_{y}$.

As in the case with the results obtained in the last section, the chiral edge states appear in the gapped phase with nonzero Chern number (Fig.~\ref{fig: Chern-edge-xOBC}\,(c-d) and Fig.~\ref{fig: Chern-edge-xOBC-eigenvector}\,(a)), whereas no edge states appear in the gapped phase with zero Chern number (Fig.~\ref{fig: Chern-edge-xOBC}\,(a-b)). However, the energy dispersions are drastically different from those with periodic boundaries (Fig.~\ref{fig: Chern-band})~\cite{remark-BC, remark-yOBC}. In particular, the spectra are entirely real in some cases (Fig.~\ref{fig: Chern-edge-xOBC}\,(b, f)), which are to be contrasted with the corresponding periodic cases (Fig.~\ref{fig: Chern-band}\,(b, f)). The eigenstates also differ from the periodic case: all the eigenstates are localized at either left or right edge (Fig.~\ref{fig: Chern-edge-xOBC-eigenvector}). The energy dispersions of the chiral edge states are obtained as
\begin{equation} \begin{split}
E_{\rm left} \left( k_{y} \right)
= t \sin k_{y},~~
E_{\rm right} \left( k_{y} \right)
= - t \sin k_{y},
	\label{eq: chiral edge - energy - x}
\end{split} \end{equation}
and the corresponding edge states are given by
\begin{equation} \begin{split}
\hat{\Psi}_{\rm left} \left( k_{y} \right)
&\propto \sum_{x=1}^{L_{x}} \left( - \cos k_{y} - \frac{m-\gamma}{t} \right)^{x-1} \hat{c}_{x, k_{y}} \begin{pmatrix} 1 \\ 0 \end{pmatrix}, \\
\hat{\Psi}_{\rm right} \left( k_{y} \right)
&\propto \sum_{x=1}^{L_{x}} \left( - \cos k_{y} - \frac{m+\gamma}{t} \right)^{x-1} \hat{c}_{x, k_{y}} \begin{pmatrix} 0 \\ 1 \end{pmatrix}.
	\label{eq: chiral edge - state - x}
\end{split} \end{equation}
Here the imaginary parts of the edge states vanish in contrast to Eq.~(\ref{eq: chiral edge - energy}). In addition, the localization lengths depend on the non-Hermiticity $\gamma$ in contrast to Eq.~(\ref{eq: chiral edge - state}), and they are different according to which edge they are localized at; the left edge state is present for $\left| m-\gamma \right| < 2t$, while the right edge state is present for $\left| m+\gamma \right| < 2t$. 

Remarkably, the energy gap can be open (Fig.~\ref{fig: Chern-edge-xOBC}\,(e-h)) even in the gapless phases for the corresponding periodic systems (Fig.~\ref{fig: Chern-band}\,(e-h)), and there appears a pair of helical edge states that are localized only at the right edge (Fig.~\ref{fig: Chern-edge-xOBC}\,(e-f) and Fig.~\ref{fig: Chern-edge-xOBC-eigenvector}\,(b)). The emergence of such anomalous edge states localized only at one edge reminds us of those in a one-dimensional system with sublattice symmetry~\cite{Lee-16}. In another phase, two pairs of helical edge states appear, one localized at the right edge and the other localized at the left edge (Fig.~\ref{fig: Chern-edge-xOBC}\,(g-h)). These anomalous helical edge states are immune to disorder that does not close the energy gap, and hence they are topologically protected (see Appendix~\ref{sec: disorder} for details). Moreover, the finite-size effects are negligible in this system (see Appendix~\ref{sec: finite-size effect} for details). The phase diagram of this open-boundary system is numerically obtained as shown in Fig.~\ref{fig: Chern-xOBC-phase}, which is drastically different from that of the periodic-boundary system (Fig.~\ref{fig: Chern-phase}) in stark contrast to the Hermitian counterparts. Along with such sensitivity of the complex spectrum to the boundary conditions, recent work~\cite{Yao-18} proposed a generalized Chern number that precisely predicts the emergence of the chiral edge states even under the open boundary condition.

The emergence of the anomalous helical edge states is qualitatively understood as follows. Let us consider the topological phase transition from the gapped phase with $C=-1$ to that with $C=0$, through the $\nu_{0} \neq 0$ phase (Fig.~\ref{fig: Chern-phase}). In the $C=-1$ phase, the chiral edge states appear at both edges given as Eq.~(\ref{eq: chiral edge - state - x}), and all the eigenstates except for the left chiral edge state are localized at the right edge. If we approach the boundary between the $C=-1$ and $\nu_{0} \neq 0$ phases, the left chiral edge state gradually becomes delocalized, whereas the right chiral edge state does not. At the boundary between the $C=-1$ and $\nu_{0} \neq 0$ phases, the bulk gap closes (even under the open boundary condition along the $x$ direction), and the left chiral edge state is absorbed into the bulk eigenstates and shifted to the right edge. In fact, the left chiral edge state vanishes at the phase boundary $\left| m-\gamma \right| = 2t$, as described above. Inside the $\nu_{0} \neq 0$ phase, the bulk eigenstates acquire an energy gap, but the helical edge states remain gapless. At the boundary between the $\nu_{0} \neq 0$ and $C=0$ phases, the helical edge states are absorbed into the bulk eigenstates and disappear inside the $C=0$ phase.

Since the helical edge states emerge in the $\nu_{0} \neq 0$ phase, they may be related to the topological charges of exceptional points given by Eq.~(\ref{eq: EP - charge}). In addition, we point out that the helical edge states in Fig.~\ref{fig: Chern-edge-xOBC}\,(e-f) seem to bridge the two exceptional points that appear in the periodic systems (Fig.~\ref{fig: Chern-band}\,(e-f)).

%%%%%
\subsection{Open boundaries in both $x$ and $y$ directions}

\begin{figure}[t]
\centering
\includegraphics[width=84mm]{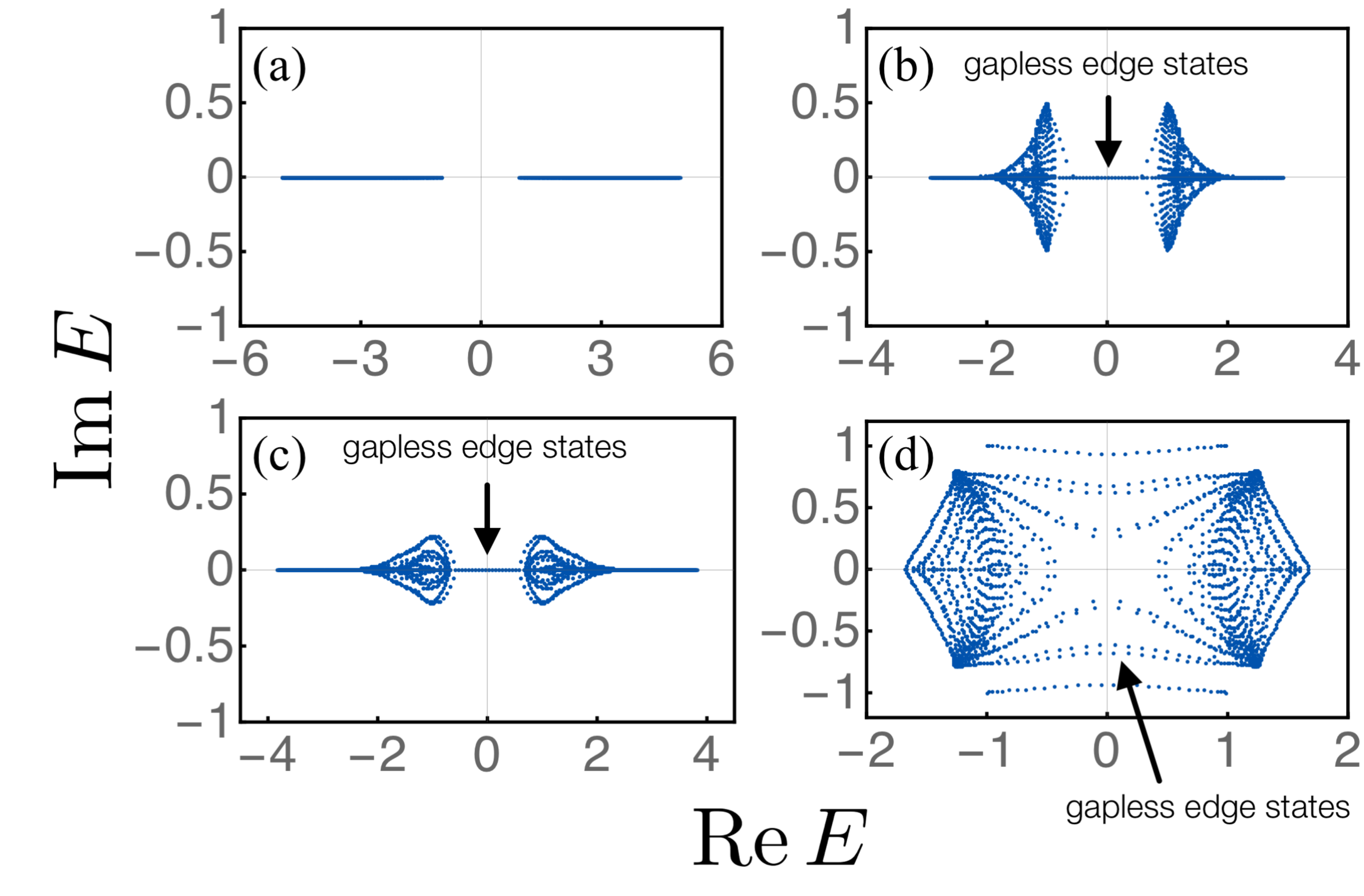} 
\caption{Complex spectrum of the non-Hermitian Chern insulator given by Eq.~(\ref{eq: Chern insulator - OBC}), which has open boundaries in both $x$ and $y$ directions ($L_{x} = L_{y} = 30$). Real and imaginary parts of the complex eigenenergies are shown for (a)~$t=1.0,\,m=-3.0,\,\gamma=0.5$, (b)~$t=1.0,\,m=-1.0,\,\gamma=0.5$, (c)~$t=1.0,\,m=-2.0,\,\gamma=1.0$, and (d)~$t=1.0,\,m=-0.2,\,\gamma=1.0$. The complex gap is open in all the cases, and the gapless edge states appear for (b-d).} 
	\label{fig: Chern-spectra-OBC}
\end{figure}

\begin{figure}[t]
\centering
\includegraphics[width=85mm]{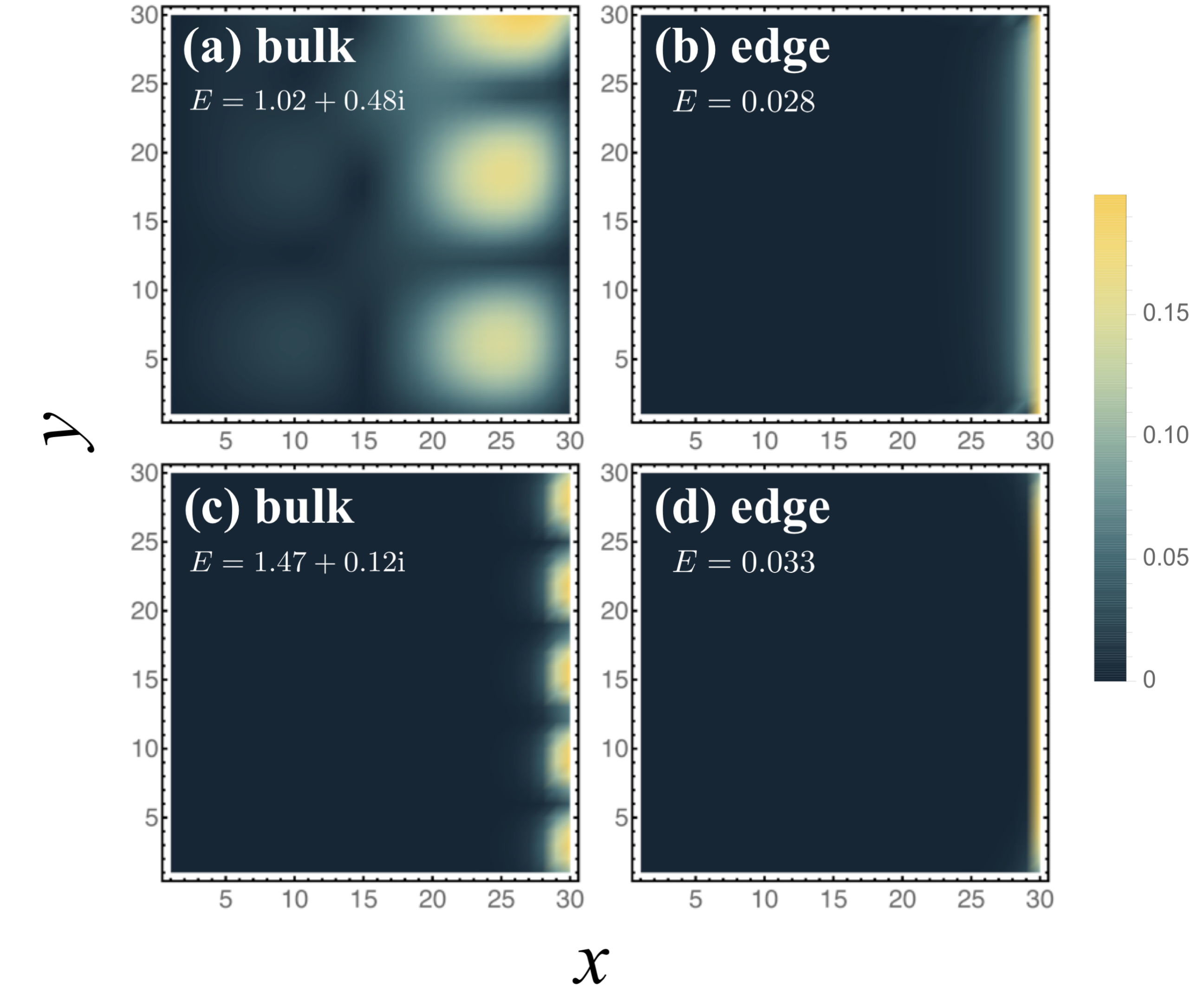} 
\caption{Wavefunctions of the non-Hermitian Chern insulator given by Eq.~(\ref{eq: Chern insulator - OBC}), which has open boundaries in both $x$ and $y$ directions ($L_{x} = L_{y} = 30$). The system is in the gapped phase that corresponds to Fig.~\ref{fig: Chern-spectra-OBC}\,(b) ($t=1.0,\,m=-1.0,\,\gamma=0.5$) for (a-b) and Fig.~\ref{fig: Chern-spectra-OBC}\,(c) ($t=1.0,\,m=-2.0,\,\gamma=1.0$) for (c-d). The gapless states (b, d) are more localized at the right edge than the bulk states (a, c).}
	\label{fig: Chern-wavefunction-OBC}
\end{figure}

Finally, we investigate the following non-Hermitian Chern insulator with open boundaries in both $x$ and $y$ directions:
\begin{equation} \begin{split}
&\hat{H} = \sum_{x} \sum_{y} \left\{ \left[ \hat{c}_{x, y+1}^{\dag} \frac{t \left( \sigma_{x} + \ii \sigma_{z} \right)}{2} \hat{c}_{x, y} + {\rm H.c.} \right] \right. \\
&~~~~~~~~~~~~~~~~~~~+ \left[ \hat{c}_{x+1, y}^{\dag} \frac{t \left( \sigma_{x} + \ii \sigma_{y} \right)}{2} \hat{c}_{x, y} + {\rm H.c.} \right] \\
&~~~~~~~~~~~~~~~~~~~~~~~~+ \left. \hat{c}_{x, y}^{\dag} \left( m\,\sigma_{x} + \ii \gamma\,\sigma_{y} \right) \hat{c}_{x, y} \right\},
		\label{eq: Chern insulator - OBC}
\end{split} \end{equation}
where $\hat{c}_{x, y}$ ($\hat{c}_{x, y}^{\dag}$) annihilates (creates) a fermion with two internal degrees of freedom on site $\left( x, y \right)$.

The numerically obtained complex spectra are shown in Fig.~\ref{fig: Chern-spectra-OBC}. Remarkably, the gapless edge states do not hybridize with each other but survive even in this open boundary condition (Fig.~\ref{fig: Chern-spectra-OBC}\,(b-d)). These gapless states are more localized than the bulk states (Fig.~\ref{fig: Chern-wavefunction-OBC}). In contrast to the Hermitian counterparts, they are localized only at the right edge and hence do not circulate around the boundary. 

%%%%% Conclusion and Outlook %%%%%
\section{Conclusion and Outlook}
	\label{sec: conclusion}

In this work, we have investigated a non-Hermitian Chern insulator and its bulk-boundary correspondence. We have found the phase diagram of the system with periodic boundaries; the gapped phases characterized by the Chern number are separated by the gapless phases characterized by the topological charges accompanied by exceptional points. We have also investigated topologically protected edge states for the same system with open boundaries. We have shown that the chiral edge states that correspond to the nontrivial topology of the bulk are robust even in this non-Hermitian system, and found the emergence of the anomalous helical edge states that are localized only at one edge. Our work serves as a two-dimensional generalization of the anomalous edge states in one dimension~\cite{Lee-16}. Such anomalous edge states may appear even in different dimensions (including three dimensions) and symmetry classes, which we leave for future work.

Another important outstanding issue is to find the topological invariants that characterize the anomalous helical edge states and to establish their bulk-boundary correspondence. In the integer quantum Hall effect, the physical origin of the chiral edge states can be attributed to the presence of a confining potential around the edges and the gauge invariance for virtual magnetic fields~\cite{Laughlin-81, Halperin-82, Buttiker-88, Hatsugai-93}. It merits further study to establish such a physical picture for the helical edge states in the non-Hermitian Chern insulator discussed in this work.

%%%%% Note %%%%%
\smallskip
{\it Note added.\,---\,}After completion of this work, we became aware of a recent related work~\cite{Kunst-18}, which also investigates a non-Hermitian Chern insulator and its edge states. Moreover, Refs.~\cite{Philip-18, Chen-18}, which appeared after the present work was submitted, study the Hall conductance of a non-Hermitian Chern insulator.

%%%%% Acknowledgement %%%%%
\section*{Acknowledgment}
K.~K. thanks Sho Higashikawa, Hosho Katsura, Tao Liu, Huan-Wen Wang, and Zhong Wang for helpful discussions. K.~K. also appreciates hospitality of Condensed Matter Theory Group at Kyoto University through the domestic junior researcher exchange program of ``Topological Materials Science," which inspired this work. This work was supported by KAKENHI Grant No.~JP18H01145 and a Grant-in-Aid for Scientific Research on Innovative Areas ``Topological Materials Science" (KAKENHI Grant No.~JP15H05855) from the Japan Society for the Promotion of Science. K.~K. was supported by the JSPS through Program for Leading Graduate Schools (ALPS). K.~S. was supported by RIKEN Special Postdoctoral Researcher Program.

%%%%% Appendix %%%%%
\appendix

\begin{figure}[t]
\centering
\includegraphics[width=86mm]{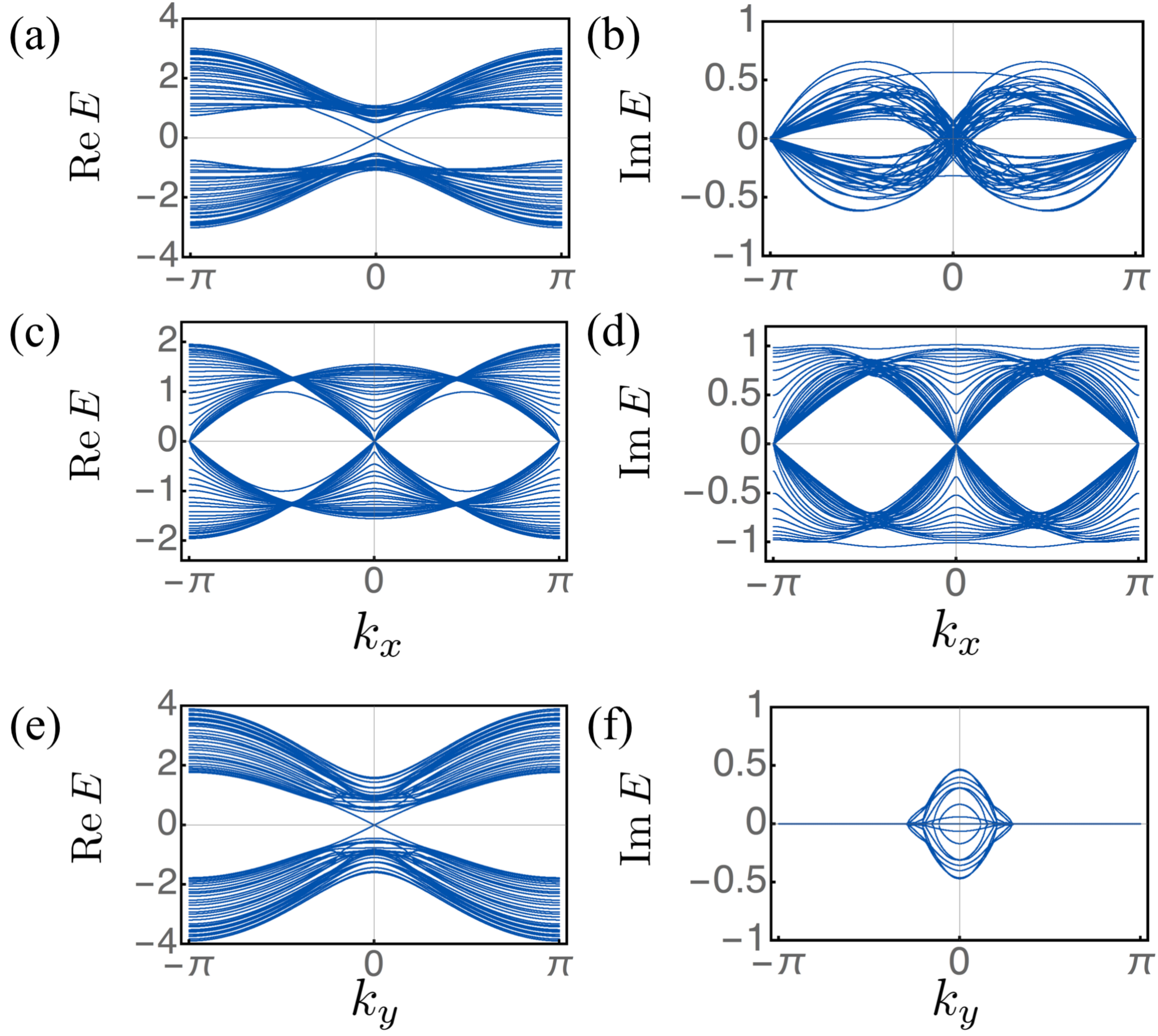} 
\caption{Stability of the chiral and helical edge states. (a-d)~Complex spectrum of the disordered non-Hermitian Chern insulator given by Eq.~(\ref{eq: Chern insulator - disorder - yOBC}) as a function of the wavenumber $k_{x}$. The periodic boundary condition is imposed in the $x$ direction, while the open boundary condition is imposed in the $y$ direction ($L_{y}=30$). The chiral edge states are robust against disorder for both (a-b)~$t=1.0$, $m_{y}=-1.0+0.5\,\epsilon_{y}$, and $\gamma_{y}=0.5+0.5\,\epsilon'_{y}$, and (c-d)~$t=1.0$, $m_{y}=-0.2+0.2\,\epsilon_{y}$, and $\gamma_{y}=1.0+0.2\,\epsilon'_{y}$, where $\epsilon_{y}$ and $\epsilon'_{y}$ are random variables uniformly distributed over $\left[ -0.5,\,0.5 \right]$. (e-f)~Complex spectrum of the disordered non-Hermitian Chern insulator given by Eq.~(\ref{eq: Chern insulator - disorder - xOBC}) as a function of the wavenumber $k_{y}$. The open boundary condition is imposed in the $x$ direction ($L_{x}=30$), while the periodic boundary condition is imposed in the $y$ direction. The helical edge states are robust against disorder for $t=1.0$, $m_{x}=-2.0+0.5\,\epsilon_{x}$, and $\gamma_{x}=1.0+0.5\,\epsilon'_{x}$.}
	\label{fig: Chern-edge-disorder}
\end{figure}

\begin{figure}[t]
\centering
\includegraphics[width=86mm]{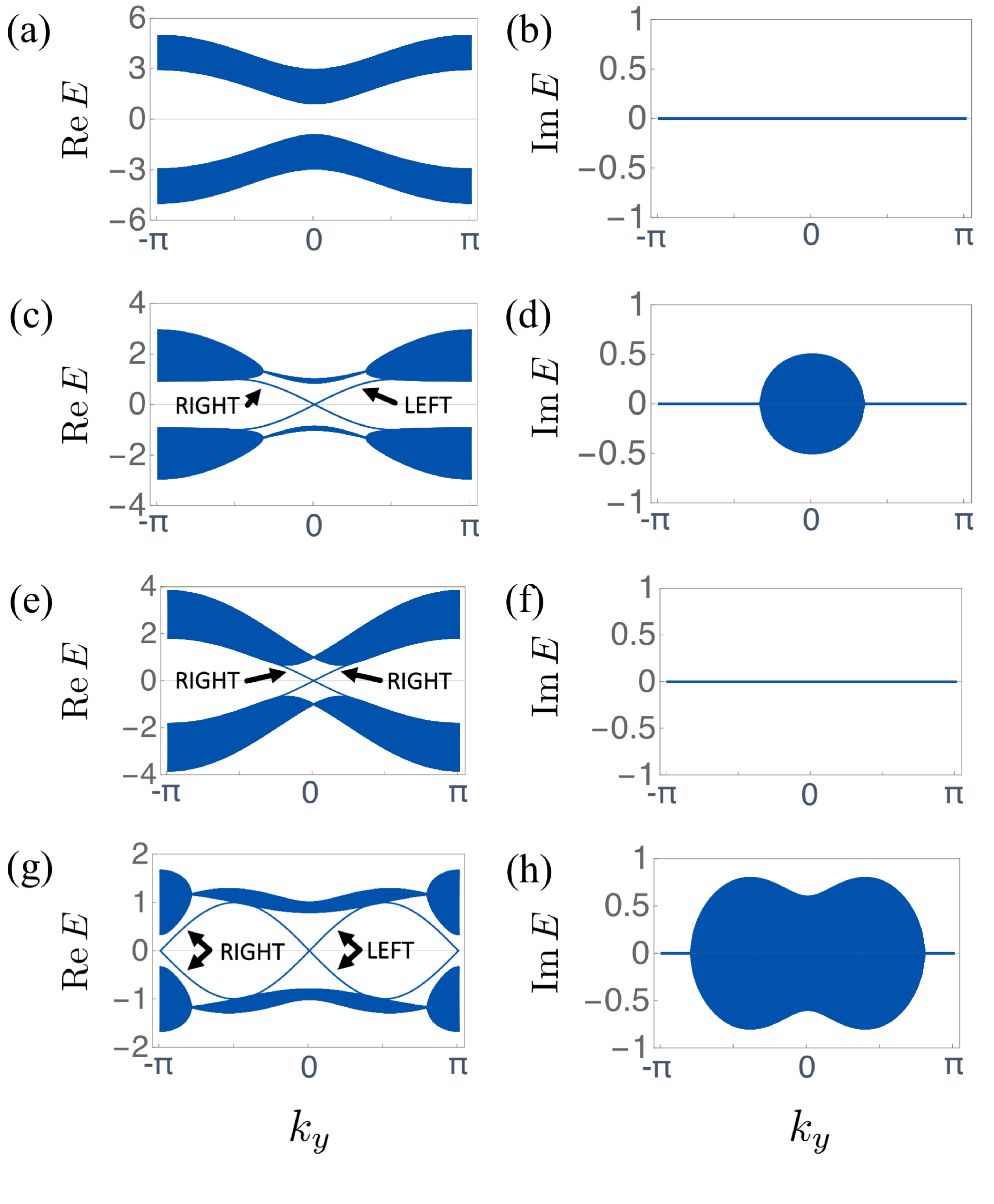} 
\caption{Complex spectrum of the non-Hermitian Chern insulator given by Eq.~(\ref{eq: Chern insulator - xOBC}), which has open boundaries in the $x$ direction ($L_{x} = 200$) and periodic boundaries in the $y$ direction. The same sets of parameters are used as in Fig.~\ref{fig: Chern-edge-xOBC}. The band gap is open and the chiral or helical edge states survive even in this larger system.}
	\label{fig: Chern-edge-xOBC-long}
\end{figure}

\section{Exact forms of the chiral edge states}
	\label{sec: chiral edge states}
	
We analytically derive the energy dispersions and wave functions of the chiral edge states in the non-Hermitian Chern insulator discussed in the main text~\cite{remark-calc}. We first consider the chiral edge state at the left edge of the Hamiltonian given by Eq.~(\ref{eq: Chern insulator - yOBC}). If the edge state is expressed as
\begin{equation}
\hat{\Psi}_{\rm left}
\propto \sum_{y=1}^{L_{y}} \lambda^{y-1} \left( \hat{c}_{k_{x}, y}\,\vec{v} \right),
\end{equation}
where $\lambda$ denotes a parameter that determines the localization length (given by $ - \left( \log \left| \lambda \right| \right)^{-1}$) and $\vec{v}$ is a two-component vector. Then the Schr\"odinger equation $[ \hat{H},\,\hat{\Psi}_{\rm left} ] = E_{\rm left}\,\hat{\Psi}_{\rm left}$ reduces to
\begin{equation}
\left( \lambda^{-1} T + M + \lambda T^{\dag} \right) \vec{v} = E_{\rm left}\,\vec{v}
	\label{eq: chiral eom - 1}
\end{equation}
in the bulk and 
\begin{equation}
\left( M + \lambda T^{\dag} \right) \vec{v} = E_{\rm left}\,\vec{v}
	\label{eq: chiral eom - 2}
\end{equation}
at the left edge. Here $T$ and $M$ are defined as $T := t \left( \sigma_{x} + \ii \sigma_{z} \right) /2$ and $M := \left( m + t \cos k_{x} \right) \sigma_{x} + \left( \ii \gamma + t \sin k_{x} \right) \sigma_{y}$, and we take the semi-infinite limit $L_{y} \to \infty$ and neglect the effect of the right edge. Combining Eqs.~(\ref{eq: chiral eom - 1}) and (\ref{eq: chiral eom - 2}), we have $T \vec{v} = 0$, which leads to $\vec{v} = \left( 1~-\ii \right)^{T}$. Hence with Eq.~(\ref{eq: chiral eom - 2}) we have
\begin{equation} \begin{split}
[ E_{\rm left} &+ \left( t \sin k_{x} + \ii \gamma \right) ] \left( 1~-\ii \right)^{T} \\
&= \left[ \lambda t + \left( m + t \cos k_{x} \right) \right] \left( -\ii~1 \right)^{T},
\end{split} \end{equation}
which gives Eqs.~(\ref{eq: chiral edge - energy}) and (\ref{eq: chiral edge - state}) in the main text. For the presence of this edge state, there exists a wavenumber $k_{x}$ that satisfies $\left| \cos k_{x} + m/t \right| < 1$, which reduces to $\left| m\right| < 2t$. The other chiral edge state at the right edge is also obtained in a similar manner. 

We next consider the chiral edge state at the left edge of the Hamiltonian given by Eq.~(\ref{eq: Chern insulator - xOBC}). If the edge state is expressed as
\begin{equation}
\hat{\Psi}_{\rm left}
\propto \sum_{x=1}^{L_{x}} \lambda_{\rm left}^{x-1} \left( \hat{c}_{x, k_{y}}\,\vec{v} \right),
\end{equation}
the Schr\"odinger equation $[ \hat{H},\,\hat{\Psi}_{\rm left} ] = E_{\rm left}\,\hat{\Psi}_{\rm left}$ reduces to
Eqs.~(\ref{eq: chiral eom - 1}) and (\ref{eq: chiral eom - 2}) with $T := t \left( \sigma_{x} + \ii \sigma_{y} \right) /2$ and $M := \left( m + t \cos k_{y} \right) \sigma_{x} + \ii \gamma\,\sigma_{y} + \left( t \sin k_{y} \right) \sigma_{z}$. We again have $T \vec{v} = 0$ and hence $\vec{v} = \left( 1~0 \right)^{T}$. With Eq.~(\ref{eq: chiral eom - 2}), we have
\begin{equation} \begin{split}
[ E_{\rm left} &- t \sin k_{y} ] \left( 1~0 \right)^{T} \\
&= \left[ \lambda_{\rm left} t + \left( m + t \cos k_{x} - \gamma \right) \right] \left( 0~1 \right)^{T},
\end{split} \end{equation}
which gives Eqs.~(\ref{eq: chiral edge - energy - x}) and (\ref{eq: chiral edge - state - x}) in the main text. While the other chiral edge state at the right edge is also obtained in a similar manner, their localizations depend differently on the non-Hermiticity $\gamma$; the left edge state appears for $\left| m-\gamma \right| < 2t$, while the right one appears for $\left| m+\gamma \right| < 2t$.

%%%%%
\section{Robustness against disorder}
	\label{sec: disorder}

The chiral and helical edge states revealed in the main text are immune to disorder. To confirm this, we first consider the following disordered non-Hermitian Chern insulator with periodic boundaries in the $x$ direction and open boundaries in the $y$ direction:
\begin{equation} \begin{split}
&\hat{H} = \sum_{k_{x}} \sum_{y} \left\{ \left[ \hat{c}_{k_{x}, y+1}^{\dag} \frac{t \left( \sigma_{x} + \ii \sigma_{z} \right)}{2} \hat{c}_{k_{x}, y} + {\rm H.c.} \right] \right. \\
&+ \left. \hat{c}_{k_{x}, y}^{\dag} \left[ \left( m_{y} + t \cos k_{x} \right) \sigma_{x} + \left( \ii \gamma_{y} + t \sin k_{x} \right) \sigma_{y} \right] \hat{c}_{k_{x}, y} \right\},
		\label{eq: Chern insulator - disorder - yOBC}
\end{split} \end{equation}
where $m_{y}$ and $\gamma_{y}$ denote the parameters characterizing disorder on site $y$. The chiral edge states appear even in the presence of disorder and hence are topologically protected (Fig.~\ref{fig: Chern-edge-disorder}\,(a-d)). We also consider the following disordered non-Hermitian Chern insulator with open boundaries in the $x$ direction and periodic boundaries in the $y$ direction:
\begin{equation} \begin{split}
&\hat{H} = \sum_{x} \sum_{k_{y}} \left\{ \left[ \hat{c}_{x+1, k_{y}}^{\dag} \frac{t \left( \sigma_{x} + \ii \sigma_{y} \right)}{2} \hat{c}_{x, k_{y}} + {\rm H.c.} \right] \right. \\
&+ \left. \hat{c}_{x, k_{y}}^{\dag} \left[ \left( m_{x} + t \cos k_{y} \right) \sigma_{x} + \ii \gamma_{x}\,\sigma_{y} + \left( t \sin k_{y}\right) \sigma_{z} \right] \hat{c}_{x, k_{y}} \right\},
	\label{eq: Chern insulator - disorder - xOBC}
\end{split} \end{equation}
where $m_{x}$ and $\gamma_{x}$ denote the parameters characterizing disorder on site $x$. As well as the chiral edge states, the helical edge states are topologically protected (Fig.~\ref{fig: Chern-edge-disorder}\,(e-f)).

%%%%%
\section{Finite-size effect}
	\label{sec: finite-size effect}

Figure \ref{fig: Chern-edge-xOBC-long} represents the complex spectrum of the non-Hermitian Chern insulator with open boundaries in the $x$ direction and periodic boundaries in the $y$ direction. In contrast to Fig.~\ref{fig: Chern-edge-xOBC} with $L_{x} = 30$, Fig.~\ref{fig: Chern-edge-xOBC-long} is on the larger systems with $L_{x} = 200$. The band gap is open and the chiral or helical edge states survive even in this larger system; the finite-size effects are thus negligible.

%%%%% Reference %%%%%


\begin{thebibliography}{}
\bibitem{remark-calc}
A systematic way to analytically obtain the edge states is described in Refs.~\cite{Mong-11, Viola-16, Kawabata-17-Majorana}.
\bibitem{remark-yOBC}
We observe that such qualitative change of the complex spectrum does not occur for the system with open boundaries only in the $y$ direction. In addition, all the eigenstates except the chiral edge states are delocalized in the bulk in this case.
\end{thebibliography}
\end{document}